\documentstyle[aps,prl,multicol,epsf]{revtex}

\newcommand{\bleq}{\ifpreprintsty
                   \else
                   \end{multicols}\vspace*{-3.5ex}{\tiny
                   \noindent\begin{tabular}[t]{c|}
                   \parbox{0.493\hsize}{~} \\ \hline \end{tabular}}
                   \fi}
\newcommand{\eleq}{\ifpreprintsty
                   \else
                   {\tiny\hspace*{\fill}\begin{tabular}[t]{|c}\hline
                    \parbox{0.49\hsize}{~} \\
                    \end{tabular}}\vspace*{-2.5ex}\begin{multicols}{2}
                    \fi}
\newcommand{\bcols}{\ifpreprintsty\else\begin{multicols}{2}\fi}
\newcommand{\ecols}{\ifpreprintsty\else\end{multicols}\fi}

\begin{document}
\bibliographystyle{prsty}
\title{On the Nature of Infrared Singularities in $d\leq 2$ Disordered 
Interacting Systems.}
  
\draft

\author{Anton Andreev$^a$, and Alex Kamenev$^b$}
\address{$^a$Institute for Theoretical Physics, 
         $^b$Department of Physics, 
 University of California Santa Barbara, CA 93106-4030.
  \\
  {}~{\rm (\today)}~
  \medskip \\
  \parbox{14cm} 
    {\rm We address the problem of  infrared singularities in the 
    perturbation theory for disordered interacting systems in $d\leq 2$. 
    We show that a typical, sufficiently large interacting system exhibits 
    a linear instability in the spin triplet channel. In the density--density 
    channel, although stability is preserved, a  large number of soft 
    modes is accumulated. These phenomena are responsible for 
    the instability of the weak--interacting fixed point. Although  
    generic, the unstable direction and soft modes are highly sample specific 
    and can not be effectively captured by conventional techniques based on an 
    averaging procedure. Rather, the instability is determined by the largest
    eigenvalues of the polarization operator. We propose to employ the 
    optimal fluctuation method for evaluating the probability of 
    such events. 
    }\bigskip \\ }

\maketitle

\bcols
The  physics of two--dimensional ($d=2$) disordered 
electronic systems has received a new impetus since the accumulation of 
experimental evidence of a metal--insulator transition in zero 
magnetic field \cite{Kravchenko94}. Other remarkable experimental 
developments are associated with $d=2$ finite size structures -- quantum dots 
\cite{Marcus96}. The data indicate the crucial role of  {\em e--e} 
interactions in these disordered (chaotic) systems. We also point out that in 
both cases spin degrees of freedom seem to play a major  role.   

On the theoretical side it was known since the work of Altshuler and Aronov 
\cite{Altshuler79} 
that various physical observables acquire infrared singular 
corrections
in the presence of {\em e--e} interactions. Using a finite system size, $L$, as 
a cutoff \cite{foot1}, 
one may write the correction to an observable, $\chi$ (which may be 
specific heat, conductivity, spin polarization etc.) as  
\begin{equation}
\frac{\delta\chi}{\chi}\propto 
\frac{ L^{2-d} } {\nu_d D} \, ,           \label{AA}
\end{equation}
where $\nu_d$ is a $d$--dimensional density of states at a Fermi level and 
$D$ is a diffusion constant. Hereafter, $L^{2-d}$ should be understood as 
$\ln L/l$ with $l$ being a microscopic length (mean free path \cite{Rudin97})
for the  $d=2$ case. 
Once the system size becomes comparable with the characteristic length 
$\xi_d\equiv (\nu_d D)^{1/(2-d)}$ ($\xi_2=le^{\nu_2 D}$),  
the validity of the first order 
perturbation theory breaks down, and expression (\ref{AA}) can not be applied. 
Despite  the fact that $\xi_d$ has an interpretation as the localization 
length, Eq.~(\ref{AA}) has nothing to do with the actual Anderson 
localization. In fact, in the presence of a small magnetic field the physical 
localization length may be much larger than $\xi_d$. Thus, the behavior 
of an observable $\chi$ at $L > \xi_d$ may be studied
quite independently of the onset of the Anderson localization. 

Finkel'stein \cite{Finkelstein83,Finkelstein84} 
has developed a  renormalization group 
approach to the problem. His results show  that a weak--disorder 
and weak--interaction fixed point is unstable at $d\leq 2$. As a result, the 
large scale behavior of the system is governed by some other fixed point. 
He has also noticed  that the system acquires 
a tendency towards a partial spin polarization and pointed out that 
the spin susceptibility diverges independently  of all (and before) other 
quantities \cite{Finkelstein84}.     
In spite of considerable further efforts in this direction 
\cite{Belitz94,Castellani86,Sachdev95}, 
the physical nature of the new fixed point 
remains to a large extent unclear. The ultimate fate of the corrections, 
Eq.~(\ref{AA}), in the region $L > \xi_d$ is also unknown. 

The purpose of this letter is to shed some light on the above questions. 
We concentrate primarily on the spin triplet interaction 
channel, restricting ourselves to few remarks concerning spin singlet 
(density--density) channel. We argue that for each {\em particular 
realization} of disorder and $L > \xi_d$ the system exhibits 
a {\em linear response} instability in the spin triplet channel. 
On the  mean--field level this instability leads to the spontaneous 
breaking of the spin--rotation symmetry 
and partial spin polarization. The questions concerning 
stability of true long-range order with respect to 
fluctuations in the thermodynamic limit are well beyond 
the  scope of this letter. For a finite size system at sufficiently low
temperatures, however, the fluctuations will not destroy spontaneous
magnetization. The existence of an unstable direction 
(in the functional space of possible spin polarizations) is a 
generic property  of any disordered system. Its concrete shape, 
however, depends on the polarization operator for a 
given disorder realization and is extremely sample dependent. 
Methods which employ an ensemble averaging procedure at 
an early stage of the calculations are bound to loose information about 
this instability since they use the disorder-averaged polarization operator. 
Instead of being linear, the effect appears to be 
encoded  into higher non--linear corrections. 
The stability of the paramagnetic state for a given system 
has very little to do with the averaged susceptibility.
The importance of fluctuations 
of various susceptibilities in disordered metals was stressed 
by several authors \cite{Spivak91,Sachdev89} (see also \cite{Spivak95} 
for applications to superconductors). 
In fact, the stability of the paramagnetic state 
is determined by atypically large eigenvalues of the polarization 
operator\cite{Sachdev89}. Thus, methods treating the optimal fluctuations
seem to be more adequate. These methods were initially developed for the 
problem of ``Lifshitz tails'' by 
Zittartz and Langer \cite{Zittartz66} and used recently in the problem 
of pre--localized states \cite{Khmelnitskii95}. Below we adopt them 
for the treatment of interacting electrons.

Consider a gas of interacting electrons, moving in a 
random potential. In terms of the slow degrees of freedom the interaction 
Hamiltonian takes the form 
\begin{equation}
H=\frac{1}{2\nu_d}\!\! \int \!\!\!\!  \int \! \!\! d^d r\, d^d r'\! \left[ 
V_s(r-r')n(r)n(r') -  \hat V_t {\bf s}(r) {\bf s}(r) \right] , 
                                                        \label{H}
\end{equation} 
where the charge, 
$n(r)=\sum\limits_{\sigma}\psi_{\sigma}^{\dagger}(r)\psi_{\sigma}(r)$, and
spin, 
${\bf s}(r)=\sum\limits_{\sigma\sigma'} 
\psi_{\sigma}^{\dagger}(r){\bf \sigma}_{\sigma\sigma'}\psi_{\sigma'}(r)$, 
densities are slowly varying on the scale $\lambda_F$;
$\hat V_t=V_t\delta(r-r')$. We write the partition 
function as an imaginary time functional integral \cite{Negele} 
and perform the Hubbard--Stratonovich transformation of the 
singlet and triplet interaction terms  
by means of fields $\Phi$ and ${\bf H}$ correspondingly. 
Having in mind developing a Landau--Ginzburg 
mean--field theory at finite temperature $T$, 
we restrict ourselves to the zero Matsubara frequency only, 
${\bf H}(r)={\bf H}(r,\omega_m=0)$. The subsequent 
integration over the fermionic degrees of freedom results in 
$\mbox{det}[1+G(i\Phi + {\bf H \sigma})]$, which may be now expanded in 
powers of $\Phi$ and ${\bf H}$. In this way one obtains 
\begin{equation}
\int \!\!\! D{\bf H}\exp\! \left\{\!\!  
-\frac{\nu_d T}{2}\!\! \int\!\!\!\! \int \!\! {\bf H}(r)\!\!
\left[ \hat V_t^{-1}\!- \hat \Pi(r,r') \right]\! {\bf H}(r')
\!+\! \hat\Gamma{\bf H}^4  \right\},
                                                        \label{FI}
\end{equation}  
where $\hat \Gamma {\bf H}^4$ designate non--linear terms. 
The central quantity of interest is the static 
polarization operator (PO), defined as 
\begin{equation}
\hat \Pi(r,r')=-\nu_d^{-1} T\sum\limits_{\epsilon_n}
G_{\epsilon_n}(r,r') G_{\epsilon_n}(r',r) \, ,        \label{P}
\end{equation} 
where $G_{\epsilon_n}(r,r')$ is a {\em sample specific} Green function and 
$\epsilon_n=\pi T(2n+1)$. 
The Gaussian part of the functional integral, Eq.~(\ref{FI}), 
becomes unstable if the positively defined Hermitian operator 
$\hat \Pi(r,r')$ has at least one eigenvalue larger than $V_t^{-1}$. 
We are faced, thus,  with the spectral problem for the PO: 
\begin{equation}
\int\!  d^dr'\, \hat \Pi(r,r') \Phi_n(r')= \lambda_n \Phi_n(r)\, ;   \label{S}
\end{equation}     
$\int|\Phi|^2d^dr=1$. Specifically, we are interested in the large 
$\lambda$ tail of the PO spectral density since the spontaneous 
symmetry breaking will be determined by the largest eigenvalue.
The spectral density is defined as 
\begin{equation}
N(\lambda)\equiv 
\left\langle 
\sum\limits_n \delta(\lambda-\lambda_n) 
\right\rangle\, ,  \label{N}
\end{equation}  
where the angular brackets denote disorder averaging. 
Employing the optimal fluctuation we shall demonstrate that 
\begin{equation}
N\left( \lambda > \lambda^{\mbox{typical}}  \right) \propto 
\exp \left\{-f\left(\lambda\frac{\nu_d D}{L^{2-d}} \right)\right\}\, ,  
                                                              \label{N1}
\end{equation} 
with $f(x)$ being a certain universal function ($\sim x^2$ -- in a simplest 
approximation). Eq.~(\ref{N1}) demonstrates that the condition for the 
appearance of a large ($\lambda >V_t^{-1}$) eigenvalue is 
$L^{2-d} > \nu_d D/V_t$ which coincides precisely with the breakdown of the 
perturbation theory, Eq.~(\ref{AA}). It is clear now from Eq.~(\ref{FI}) that 
once such an eigenvalue is formed, the system develops a non--trivial 
saddle point with ${\bf H}(r)\neq 0$, which describes a spin polarized state. 
A new expansion arround this sample--specific minimum should be developed. 
 
The (replicated) ensemble averaged version of the effective action in 
Eq.~(\ref{FI})  is 
\begin{equation}
\frac{\nu_d T}{2}\!\! \int\!\!\!\! \int \!\! {\bf H}\!
\left[ \hat V_t^{-1}- \langle \hat \Pi \rangle \right] \!{\bf H} +
\left( 
\langle \hat \Gamma\rangle - \frac{1}{4}
\langle\langle \hat \Pi^2 \rangle\rangle \right) {\bf H}^4 +\ldots  \, .
                                                        \label{AE}
\end{equation}  
The average PO,  
$\langle  \Pi(r,r')\rangle =\delta(r-r')$, is an operator with all 
eigenvalues equal to unity and thus with no tails in spectral density 
\cite{foot3}.  This fact ensures the stability of 
the Gaussian integral for $V_t^{-1}>1$, which is the Stoner criterion 
for the ferromagnetic instability \cite{Doniach}. The information 
about anomalously large 
eigenvalues  of $\hat \Pi$ is now hidden in the higher  order non--linear 
terms. Thus the basic fact about the presence of a linear instability 
in the theory appears to be obscured by the averaging procedure. 

Let us make a few remarks concerning the singlet interaction channel. The 
corresponding functional integral over a scalar field $\Phi$ has a Gaussian 
part of the form 
\begin{equation}
\frac{\nu_d T}{2}\! \int\!\!\!\int d^dr\,  d^dr' \Phi(r)
\left[ V_s^{-1}(r-r') + \hat \Pi(r,r') \right] \Phi(r')\, .
                                                        \label{SC}
\end{equation}  
Note the plus sign in front of PO! The kernel is now strictly positively 
defined and can not exhibit an instability. It may have, however, very soft 
(and long-wavelength) modes associated with a {\em small} $\lambda$ 
($\lambda \geq 0$) tail of the spectral density $N(\lambda)$ \cite{foot2}.
These atypically soft modes may lead to  singularities in the singlet channel. 
One should be careful with these arguments, however, since there is no 
reason to neglect non--zero Matsubara frequencies in this situation.  
For $\omega_m > DQ^2$ even the average PO has small eigenvalues. 
The fact that the divergence of singlet and 
triplet quantities is associated with the opposite tails of 
$N(\lambda)$ may be related to Finkel'stein's results quoted above.

We turn now to the evaluation of the tails of the PO spectral density,  
$N(\lambda)$. To this end we define the corresponding Green function operator 
as
\begin{equation}
\hat {\cal G}_{\lambda}(r,r') = (\hat \lambda -i\eta -\hat \Pi)^{-1}= 
\sum\limits_n \frac{\Phi_n(r)\Phi_n^*(r')}{\lambda-i\eta-\lambda_n}\, 
                                                        \label{GF}
\end{equation}  
and 
$N(\lambda)=\pi^{-1} \Im \mbox{Tr} \langle \hat {\cal G}_{\lambda} \rangle$. 
To perform disorder averaging we need to know the statistics of the random 
operator  $\hat \Pi(r,r')$. A straightforward diagrammatic calculation 
of  its second cumulant  yields at $T=0$ 
\begin{eqnarray}
&&Q^{rr'}_{ss'}\equiv  
\langle\langle \hat \Pi(r,r')\hat \Pi(s',s)\rangle\rangle =  
                                                        \label{Q0} \\
&&\frac{1}{2\pi^2\nu_d^2}
\int\limits_0^{\infty}\!  \varepsilon d \varepsilon
D_{\varepsilon}(r-s')D_{\varepsilon}(s'-r')D_{\varepsilon}(r'-s)
D_{\varepsilon}(s-r) \nonumber \, . 
\end{eqnarray}
Here $D_{\varepsilon}(r)$ is a propagator of the classical diffusion 
operator 
\begin{equation}
D_{\varepsilon}(r)=\sum\limits_Q 
\frac{e^{iQr} }{DQ^2 +|\varepsilon|} = 
\frac{L_{\varepsilon}^{2-d} }{D} F_d\left(\frac{r}{L_{\varepsilon}}\right) ,
                                                        \label{D}
\end{equation}  
where $L_{\varepsilon}=\sqrt{D/\varepsilon}$ and $F_d(x)$ is a dimensionless 
function with the assimptotic behavior $F_d(0)=\mbox{const}$ and 
$F_d(x\rightarrow\infty)\sim e^{-x}$. Performing the energy integration in 
Eq.~(\ref{Q0}) one obtains 
\begin{equation}
Q^{rr'}_{ss'}\approx \frac {(\nu_d D)^{-2}} 
{\left( |r-s'|+|s'-r'|+|r'-s|+|s-r| \right)^{4(d-1)} }\, .
                                                        \label{Q1}
\end{equation}  
We can perform now the averaging of the Green function,  
$\hat {\cal G}_{\lambda}$, assuming the Gaussian distribution for $\hat \Pi$. 
This is certainly not an exact procedure because the higher order cumulants 
of $\hat \Pi$ are not negligible in the most interesting parameter region 
$L>\xi_d$. { We shall follow, however, this idea   since it allows us to 
illustrate the method and provides a basis for further generalizations. 
We thus obtain }
\begin{equation}
\left\langle \hat {\cal G}_{\lambda}(R,R')\right\rangle = 
\int\!\!\! D\hat \Pi\, 
e^{-\frac{1}{2}\hat \Pi_{rr'}(Q^{-1})^{rr'}_{ss'}\hat\Pi^{s's} +
\ln \hat {\cal G}_{\lambda}(R,R') } \, , 
                                                        \label{AGF}
\end{equation}   
where integration over  repeated coordinates is implicit. 
The average value of the PO is omitted in this expression, since it leads to 
a trivial redefinition $\lambda\rightarrow \lambda-1$ only. We argue now, 
following Zittartz  and Langer \cite{Zittartz66}, that large $\lambda$ tails 
are determined by the saddle point of this functional integral. Variation 
over $\hat \Pi_{rr'}$ leads to the equation for the optimal realization of 
the PO  
\begin{equation}
\bar \Pi(r,r')= Q^{rr'}_{ss'} 
\frac{\hat {\cal G}_{\lambda}(R,s) \hat {\cal G}_{\lambda}(s',R')} 
{\hat {\cal G}_{\lambda}(R,R')}  \, . 
                                                        \label{OP}
\end{equation} 
In the close vicinity of some atypically large  eigenvalue, $\lambda_0$,  
the Green 
function may be well approximated by the single term in the 
sum, Eq.~(\ref{GF}),  
$\hat {\cal G}_{\lambda}(r,r') = \Phi_0(r)\Phi_0^*(r')/(\lambda-\lambda_0)$, 
leading to 
\begin{equation}
\bar \Pi(r,r')= \frac{1} {\lambda-\lambda_0}\,  
Q^{rr'}_{ss'}  \Phi_0^*(s)\Phi_0(s') \,.          
                               \label{OP1}
\end{equation}
Again the $d$--dimensional integration is assumed 
over  repeated coordinates.
Finally,  substitution of Eq.~(\ref{OP1}) into Eq.~(\ref{S}) results in the 
nonlinear self-consistency equation for the typical eigenfunction corresponding 
to a large eigenvalue, $\lambda_0$,  
\begin{equation}
\lambda_0 \Phi_0(r) = \frac{1} {\lambda-\lambda_0}\,  
Q^{rr'}_{ss'}  \Phi_0(r')\Phi_0^*(s)\Phi_0(s') \, .                       
                                                     \label{SCE}
\end{equation} 
To avoid the solution of this non--linear integral equation one may use  
purely
dimensional arguments. To this end let us write this equation in terms of 
dimensionless coordinates $x=r/L$ and dimensionless eigenfunctions 
$\phi_0(x)=L^{d/2} \Phi_0(r)$. We also employ Eq.~(\ref{Q1}) for the cumulant 
$Q$:
\begin{eqnarray}
\label{SCE1}
&& \phi_0(x) = 
\left[ \frac{L^{2-d} } {\nu_d D\sqrt{\lambda_0(\lambda-\lambda_0)} }\right]^2 
\times \\
&&\int\limits_0^1  \!\! 
\frac{d^dx'\, d^dy\, d^dy'\, \phi_0(x')\phi_0^*(y)\phi_0(y')}
{(|x-y'|+|y'-x'|+|x'-y|+|y-x|)^{4(d-1)} } \nonumber \, .              
\end{eqnarray}  
To ensure the existence of a normalizable 
$\left(\int\limits_0^1|\phi_0|^2 d^dx =1\right)$ 
solution of this equation one must 
require that the expression in the square brackets on its r.h.s. is of order 
unity. We conclude that 
\begin{equation}
\lambda-\lambda_0 \approx \lambda_0^{-1} 
\left(  \frac{L^{2-d} } {\nu_d D } \right)^2\, .                    
                                                     \label{LAM}
\end{equation} 
The last step is to find the statistical weight of the optimal realization 
$\bar\Pi$, which has $\lambda_0$ as an eigenvalue. Its statistical weight 
is obviously given by 
$\bar\Pi Q^{-1} \bar\Pi=  \lambda_0/(\lambda-\lambda_0)$, where we 
employed  Eqs.~(\ref{OP1}), (\ref{SCE}) and normalization condition for 
$\Phi_0$. The last expression together with Eq.~(\ref{LAM}) implies that 
\begin{equation}
N(\lambda_0)\propto 
\exp \left\{-c_d\left(\lambda_0\frac{\nu_d D}{L^{2-d}} 
\right)^2 \right\}\, ,   
                                                              \label{N2}
\end{equation}  
where $c_d$ is a numerical constant of order unity. 
The fact that Gaussian tails are obtained may be traced back to the 
assumption about the Gaussian distribution of the PO and should not be taken 
too seriously in the regime $L>\xi_d$. However, the dependence of the spectral 
density on the parameter $\lambda \nu_d D/L^{2-d}$ follows from pure 
dimensional analysis and may be justified for more realistic assumptions 
concerning  PO statistics. In fact, the Gaussian spectral density,  
Eq.~(\ref{N2}), is the most pessimistic estimate for the tails. Indeed, all 
the effects we have neglected (higher order cumulants of $\hat \Pi$, and 
onset of Anderson localization) should increase fluctuations of the PO, 
making the tails decrease slower than Gaussian. It remains a challenging 
problem to calculate the spectral density of the PO under more realistic 
assumptions. The ``pessimistic'' result, Eq.~(\ref{N2}), is already sufficient 
to demonstrate our main point: one may find an arbitrarily large eigenvalue 
of the PO if the system size is taken to be large enough. This leads to a 
ferromagnetic  instability in the triplet channel for arbitrarily small 
interaction, $V_t$. 

In the most interesting $d=2$ case the argument in  Eq.~(\ref{N2}) has 
the form 
$\lambda_0 g (\ln L/l)^{-1}$, where $g=\nu_2 D >1$ is a dimensionless 
conductance of a $2d$ metal. To find an eigenvalue of the PO  
$\lambda_0 > V_t^{-1}>1$ 
one should typically consider a sample with a size, $L>l\exp\{g/V_t\}$. 
This still may be much smaller than the localization length 
in the unitary ensemble,
which is of the order $l\exp\{g^2\}$. This is to say that, 
contrary to a naive 
interpretation of  Eq.~(\ref{N2}),  one may encounter the ferromagnetic  
instability well inside the metallic side of the Anderson transition.  

We want to point out that in low dimensional ($d\leq2$) systems 
the spontaneous symmetry breaking occures into an {\em extended} 
state,
rather than into a localized one (as it happens in 3d 
systems~\cite{Sachdev89}). This is clear from the extended 
nature the solution of Eq.~(\ref{SCE1}).
In this case the role of nonlinear terms ( $\Gamma {\bf H}^4$ in 
Eq.~(\ref{FI}) ) reduces to merely determining the amplitude 
of the spontaneous globally coherent magnetization.

The aim of this letter is to elucidate the source of some problems in 
the theory of $d\leq 2$ disordered interacting electrons. The traditional  
perturbation theory and RG treatment lead to infrared divergence and 
instability of the weak--coupling fixed point correspondingly. They do not 
provide a simple physical reason for the singularities, nor do they explain 
the nature of the strong coupling fixed point. 
We argue that at least some of the difficulties can be traced to 
(i) the ferromagnetic instability of a given system in the triplet channel, 
and (ii) to the accumulation of many soft modes in the singlet channel. 
The latter necessitates taking into account  non--linear  
terms in the action. Both the  unstable directions and the soft modes, 
although generic, are sample specific and do not survive 
(traditional) ensemble averaging procedure. It would be desirable 
to construct a theory, which first adjusts the integration directions 
(bosonic fields) to the concrete sample configuration and only then 
performs the averaging.

Discussions with I.~Aleiner, A.~Altland, B.~L.~Altshuler, 
R.~N.~Bhatt, I.~V.~Lerner, S. Sachdev and B.~Simons  
are highly acknowledged. This research 
was supported by the NSF Grant No.~PHY 94-07194.
A.K. was partially supported by the Rothschild fellowship.

\ecols

\begin{thebibliography}{10}
\bibitem{foot1}
Alternately one may think about infinite system and use a temperature 
length, $L_T\equiv\sqrt{D/T}$, as cutoff.
\bibitem{foot3}
In fact, $\langle \hat \Pi \rangle (r-r')$ has smaller eigenvalues associated 
with eigenfunctions oscillating at $2k_F$, which are not important in the 
present context.
\end{thebibliography}
\end{document}